\documentclass[journal]{IEEEtran}
\usepackage{setspace}
\ifCLASSINFOpdf
\else
\fi
\usepackage[cmex10]{amsmath}

 \usepackage[hyperindex=true,pdftitle={},
pdfauthor={St\'ephane Guerrier},colorlinks=TRUE,
pagebackref=false,citecolor=blue,plainpages=false,
pdfpagelabels]{hyperref} 

\usepackage{verbatim} 
\usepackage{amsthm}
\usepackage{amsbsy}
\usepackage{amssymb}
\usepackage{amsbsy}
\usepackage{amsfonts}
\usepackage{graphics}
\usepackage[pdftex]{graphicx}
\usepackage{bm}
\usepackage{bbm}
\usepackage[usenames]{color}
\usepackage{mathtools}
\usepackage{authblk}
\usepackage{booktabs}
\usepackage{nicefrac}
\usepackage{enumitem}
\usepackage{bbding}
\usepackage{soul}
\usepackage[numbers]{natbib}
\usepackage{nicefrac}
\usepackage{placeins}
\usepackage{xcolor}
\usepackage{siunitx}
\usepackage{tabularx,booktabs}

\usepackage{xr-hyper} 

\DeclareMathOperator*{\argmin}{argmin}
\DeclareMathOperator*{\cov}{cov}

\DeclareMathOperator*{\var}{var}
\DeclareMathOperator*{\tr}{tr}

\DeclareSymbolFont{lettersA}{U}{txmia}{m}{it}
\DeclareMathSymbol{\real}{\mathord}{lettersA}{"92}
\DeclareMathSymbol{\field}{\mathord}{lettersA}{"83}

\def\boxit#1{\vbox{\hrule\hbox{\vrule\kern3pt
          \vbox{\kern3pt#1\kern3pt}\kern3pt\vrule}\hrule}}

\def\sgcomment#1{\vskip0mm\boxit{\vskip 0mm{\color{blue}\bf#1}
		 {\color{blue}\bf -- SG\vskip 1mm}}\vskip 0mm}
\def\jjcomment#1{\vskip0mm\boxit{\vskip 0mm{\color{red}\bf#1}
		 {\color{red}\bf -- JJ\vskip 1mm}}\vskip 0mm}
\definecolor{pinegreen}{rgb}{0.0, 0.47, 0.44}

\def\hxcomment#1{\vskip0mm\boxit{\vskip 0mm{\color{green}\bf#1}
		 {\color{blue}\bf -- HX\vskip 1mm}}\vskip 0mm}

\def\mkhcomment#1{\vskip0mm\boxit{\vskip 0mm{\color{violet}\bf#1}
		 {\color{violet}\bf -- MKh\vskip 1mm}}\vskip 0mm}
\def\yzcomment#1{\vskip0mm\boxit{\vskip 0mm{\color{blue}\bf#1}
		 {\color{blue}\bf -- YZ\vskip 1mm}}\vskip 0mm}
	
\newtheoremstyle{mytheoremstyle} 
    {0.3cm}                      
    {0cm}                        
    {\itshape}                   
    {}                           
    {\scshape}                   
    {: }                          
    {0em}                       
    {}  

\theoremstyle{mytheoremstyle}
\newtheorem{Theorem}{Theorem}
\newtheorem{Lemma}{Lemma}

\newtheorem{Corollary}{Corollary}

\newtheoremstyle{myExampleRemarkstyle} 
    {0.3cm}                    
    {0cm}                           
    {\itshape}                   
    {}                           
    {\scshape}                   
    {: }                          
    {0em}                       
    {}  

\theoremstyle{myExampleRemarkstyle}
 
\newtheorem{Remark}{Remark}
\newtheorem{Assumption}{Assumption}

\newtheoremstyle{simuStyle}
{0.3cm} 
{0cm} 
{} 
{} 
{\bfseries} 
{.} 
{0em} 
{} 

\theoremstyle{simuStyle}

\newtheoremstyle{stratStyle}
{0.3cm} 
{0cm} 
{} 
{} 
{\scshape} 
{: } 
{0em} 
{} 

\theoremstyle{stratStyle}

\DeclareSymbolFont{lettersA}{U}{txmia}{m}{it}
\DeclareMathSymbol{\real}{\mathord}{lettersA}{"92}
\DeclareMathSymbol{\field}{\mathord}{lettersA}{"83}

\def\real{{\rm I\!R}}

\DeclareMathOperator*{\Int}{Int}

\def\0{{\bf 0}}

\DeclareMathOperator*{\f}{f} 
\DeclareMathOperator*{\argzero}{argzero}

\def\0{\mathbf{0}}

\newcolumntype{Y}{>{\centering\arraybackslash}X}

\begin{document}

\title{Wavelet-Based Moment-Matching Techniques\\ for Inertial Sensor Calibration}
%
%
%

\author{St\'ephane~Guerrier$^{*,\dag}$, Juan~Jurado$^*$, Mehran~Khaghani$^*$, Gaetan~Bakalli, Mucyo~Karemera, Roberto~Molinari, Samuel~Orso, John~Raquet, Christine~M.~Schubert~Kabban, Jan~Skaloud, Haotian~Xu \& Yuming~Zhang

\thanks{\textit{$^*$The first three authors contributed equally and are alphabetically ordered. $^{\dag}$indicates corresponding author.}}
\thanks{\textbf{S. Guerrier} is an Assistant Professor, Faculty of Science \& Geneva School of Economics and Management, University of Geneva, 1205, Switzerland. (E-mail: Stephane.Guerrier@unige.ch).}%
\thanks{\textbf{J. Jurado} is the Director of Education, U.S. Air Force Test Pilot School, Edwards AFB, CA, 93523, USA. (E-mail: Juan.Jurado.1@us.af.mil).} 
\thanks{\textbf{M. Khaghani} is a Postdoctoral Scholar, Geneva School of Economics and Management, University of Geneva, 1205, Switzerland. (E-mail: Mehran.Khaghani@unige.ch).} 
\thanks{\textbf{G. Bakalli} is a PhD candidate, Geneva School of Economics and Management, University of Geneva, 1205, Switzerland. (e-mail: gaetan.bakalli@unige.ch).} 
\thanks{\textbf{M. Karemera} is a Postdoctoral Scholar, Geneva School of Economics and Management, University of Geneva, 1205, Switzerland. (E-mail: mucyo.karemera@unige.ch)}
\thanks{\textbf{R. Molinari} is a Postdoctoral Scholar, Department of Statistics, Pennsylvania State University, PA, 16801, USA. (E-mail: rum415@psu.edu)}
\thanks{\textbf{S. Orso} is a Postdoctoral Scholar, Geneva School of Economics and Management, University of Geneva, 1205, Switzerland. (E-mail: Samuel.Orso@unige.ch)} 
\thanks{\textbf{J. Raquet} is the Director, IS4S-Dayton, Integrated Solutions for Systems, Inc., Beavercreek, OH, 45324, USA. (E-mail: John.Raquet@is4s.com).} 
\thanks{\textbf{C. M. Schubert Kabban} is an Associate Professor, Department of Mathematics and Statistics, Air Force Institute of Technology, OH, 45324, USA. (E-mail: Christine.Schubert@afit.edu).}
\thanks{\textbf{J. Skaloud} is Lecturer, Geodetic Engineering Laboratory, \'Ecole Polytechnique F\'ed\'erale de Lausanne, 1015, Switzerland (Email: jan.skaloud@epfl.ch).}
\thanks{\textbf{H. Xu} is a PhD candidate, Geneva School of Economics and Management, University of Geneva, 1205, Switzerland. (E-mail: Haotian.Xu@unige.ch).}
\thanks{\textbf{Y. Zhang} is a PhD candidate, Geneva School of Economics and Management, University of Geneva, 1205, Switzerland. (E-mail: Yuming.Zhang@unige.ch).}
\thanks{DISTRIBUTION STATEMENT A. Approved for public release; Distribution is unlimited 412TW-PA-19483}} 

%
%

\markboth{}%
{Shell \MakeLowercase{\textit{et al.}}: Bare Demo of IEEEtran.cls for Journals}
%



\maketitle

\begin{abstract}
The task of inertial sensor calibration has required the development of various techniques to take into account the sources of measurement error coming from such devices. The calibration of the stochastic errors of these sensors has been the focus of increasing amount of research in which the method of reference has been the so-called ``Allan variance slope method'' which, in addition to not having appropriate statistical properties, requires a subjective input which makes it prone to mistakes. To overcome this, recent research has started proposing ``automatic'' approaches where the parameters of the probabilistic models underlying the error signals are estimated by matching functions of the Allan variance or Wavelet Variance with their model-implied counterparts. However, given the increased use of such techniques, there has been no study or clear direction for practitioners on which approach is optimal for the purpose of sensor calibration. This paper formally defines the class of estimators based on this technique and puts forward theoretical and applied results that, comparing with estimators in this class, suggest the use of the Generalized Method of Wavelet Moments as an optimal choice.
\end{abstract}



\begin{IEEEkeywords}
Allan Variance, Wavelet Variance, Inertial Measurement Unit, Generalized Method of Wavelet Moments, Stochastic Error, Slope Method, Autonomous Regression Method for Allan Variance
\end{IEEEkeywords}


\section{Introduction}\label{sec:intro}
The identification of a probabilistic time series model and the estimation of its relative parameters for the error signal issued from various sensors, such as inertial sensors, is a key challenge in many fields of engineering that has led to a great deal of research being produced. Aside from the size of the calibration data which can entail computational burdens for the mentioned estimation tasks, the stochastic errors of these signals are often complex in nature since they can be characterized by composite (latent) stochastic processes where different underlying models contribute to the observed error signal. Although different approaches exist to perform estimation for the parameters of these processes, the currently adopted standard method for modeling the stochastic error of inertial sensors is the ``Allan Variance Slope Method'' (AVSM)~\cite{IEEE660628} that relies on the Allan Variance (AV) which is widely accepted as being a quantity of reference for the calibration of the stochastic errors issued from (low-cost) inertial sensors. Indeed, the AVSM relies on the fact that certain stochastic processes contributing to the overall signal (such as white noises and random walks) are identifiable based on the slope of the plot. Based on this property, the AVSM requires practitioners to (i) make a log-log plot of the empirical AV of an error signal, (ii) detect the regions of the plot which best represent an assumed model, (iii) estimate the parameters of the latter by estimating the slope of the AV within the selected region (based on which model parameters can be found). This approach is currently widely practiced in industry and academia making it a method of reference for inertial sensor calibration. Despite its popularity however, the AVSM procedure is lengthy and prone to (human) errors as well as having been proven to be statistically inconsistent~\cite{guerrier2016theoretical} thereby implying that, being among others subjective in nature, the resulting parameter estimates can be severely biased and do not improve as the length of the observed signal increases.

For the above reasons (i.e. statistical inconsistency and subjective nature of the AVSM), the literature has proposed different alternatives that either make this procedure autonomous or use the AV in a way to deliver consistent estimations (or both). Among these proposals, we can find those that make use of the linearity of logarithmic transforms of the AV to apply regression methods to estimate the parameters of the stochastic models assumed for the observed error signals. A recent example is given in~\cite{Jurado19} where a regularized regression approach is applied to the logarithm of the AV with base 10. Another approach is based on a linear transformation of the AV, more specifically the Haar Wavelet Variance (WV), where a generalized least square approach inverses the mapping between the model-implied WV and the empirical WV~\cite{guerrier2013wavelet}. These moment-matching techniques belong to the class of the generalized method of moments estimators (introduced in \cite{hansen1982large}) where the considered ``moments'' are either the AV, the WV or \textit{functions} of one of these quantities. Given the presence of different moment-matching approaches based on the WV (AV) in order to automatize the sensor calibration process, it is important to understand how these approaches compare both from a theoretical as well as a practical point of view. Indeed, it would be appropriate to study these methods so that practitioners have some criteria that would allow them to choose the approach that best suits their requirements.

For the above reasons, this paper intends to study the properties of this class of moment-matching approaches and put forward a proposed optimal moment-matching technique for inertial sensor calibration. Based on this goal, the paper is organized as follows. In Sec. \ref{sec:notation}, we summarize the notational convention used throughout the paper. In Sec. \ref{sec:gmwm} we discuss the class of moment-matching estimators based on (functions of) the WV and formally compare them. Sec. \ref{sec:simulations} compares calibration parameter estimation results using some existing moment-matching approaches based on the calibration of an accelerometer and a gyroscope from an STIM-300 Inertial Measurement Unit (IMU). Finally, in Sec. \ref{sec:conclusion}, we summarize our findings and conclusions.

\section{Notational Convention}
\label{sec:notation}

\textit{Conventions}
\vspace{0.4cm}
\begin{center}
\begin{tabular}{ l l }
 $(x_t)$ & refers to a sequence of values indexed \vspace{0.1cm}\\
 & by integer $t$ \\ 
 $x_t$ & refers to the $t$th value of a sequence \vspace{0.1cm}\\ 
 $Y_t$ & refers to a random variable indexed \\
 & by integer $t$ \vspace{0.1cm}\\
 $y_t$ & refers to a realization of $Y_t$ indexed \\
 & by integer $t$ \vspace{0.1cm}\\
 $\real_{+}$ & refers to the set of positive real numbers \vspace{0.1cm}\\
 $\real_{-}$ & refers to the set of negative real numbers \vspace{0.1cm}\\
 $\mathcal{C}^1(\mathcal{A}, \mathcal{B})$ & refers to the set of functions from the set $\mathcal{A}$ \\
 & to the set $\mathcal{B}$ whose first derivatives are\\
 & continuous \vspace{0.1cm} \\
 $||\mathbf{x}||_\mathbf{A}^2$ & denotes the squared Mahalanobis distance,\\
 & i.e. $||\mathbf{x}||_\mathbf{A}^2 := \mathbf{x}^T \mathbf{A} \mathbf{x}$ where $\mathbf{x} \in \real^q$ \\
 & and $\mathbf{A} \in \real^{q \times q}$ \vspace{0.1cm} \\
 $|| \mathbf{x} ||_2$ & denotes the {\color{black}$l_2$-norm} of vector $\mathbf{x} \in \real^q$, \\
 & i.e. $|| \mathbf{x} ||_2 := (\sum_{i=1}^q \mathbf{x}_i^2)^{1/2}$ \vspace{0.1cm}\\
 {\color{black}$\lVert\mathbf{A}\rVert_S$} & {\color{black}denotes the matrix spectral norm.}\\
\end{tabular} 
\end{center}

\vspace{1.8cm}
\textit{Important Notations}
\vspace{0.2cm}
\begin{center}
\begin{tabular}{ l l }
$\bm{\Theta}$ & the parameter space \vspace{0.1cm}\\
 $\bm{\theta}$ & $(p \times 1)$ \textit{generic} parameter vector such that \\
 & $\bm{\theta} \in \bm{\Theta} \subset \real^p$ \vspace{0.1cm} \\ 
 $\bm{\theta}_0$ & $(p \times 1)$ \textit{true} parameter vector such that \\
 & $\bm{\theta}_0 \in \bm{\Theta} \subset \real^p$ \vspace{0.1cm} \\ 
 $F_{\bm{\theta}}$ & data generating model parameterized by $\bm{\theta}$\vspace{0.1cm} \\
%
$\mathcal{J}$ & $\mathcal{J} := \left\{x \in \mathbb{N} \; | \; p \leq x < \log_2(T) \right\}$ \vspace{0.1cm}\\
$J$ & an element in the set $\mathcal{J}$, i.e. an integer denoting \\ & the the number of scales such that it is at least \\
 & the same as the number of parameters but smaller \\
 & than $\log_2{(T)}$ \vspace{0.1cm}\\
$\bm{\nu}$ & $(J \times 1)$ Wavelet variance or Allan variance vector \vspace{0.1cm}\\
 $\bm{\nu}(\bm{\theta})$ & $(J \times 1)$ Wavelet variance or Allan variance vector \\
 & implied by $\bm{\theta}$ assuming that $F_{\bm{\theta}}$ corresponds to \\ 
 & the true data generating process \vspace{0.1cm}\\
 $\mathbf{f}(\cdot)$ & a known vector-valued function such that \\
 & $\mathbf{f}: \, \real_+^J \mapsto \mathcal{G} \subset \real^J$ \vspace{0.1cm}\\
 $\bm{\Omega}$ & a positive definite matrix in $\real^{J \times J}$ \\
 $\hat{\bm{\Omega}}$ & an estimate of the matrix $\bm{\Omega}$ \vspace{0.1cm}\\
 
 $|\cdot|$ & denotes the absolute value \vspace{0.1cm}\\
 $\mathbf{A} \boxtimes \mathbf{B}$ & we have that $\mathbf{A} \boxtimes \mathbf{B} := \mathbf{A} \mathbf{B}\mathbf{A}^T$ where $\mathbf{A} \in \real^{k \times d}$ \\
 & and $\mathbf{B} \in \real^{d \times d}$ \vspace{0.1cm}\\
 $T_j$ & number of wavelet coefficients at scale $j \in \mathbb{N} \setminus \{0\}$,\\
 & $T_j := T - 2^j + 1$ \\
\end{tabular}
\end{center}

\section{Generalized Method of Wavelet Functional Moments}
\label{sec:gmwm}
In order to formalize the framework of reference for this paper, we firstly consider the time series $\left(X_t\right)_{t = 1, \ldots, T}$ which is supposedly generated by a composite stochastic process $F_{\bm{\theta}}$ delivered by the sum of independent sub-processes. We let $F_{\bm{\theta}_0}$ denote the true data-generating process, which is assumed known up to the value of $\bm{\theta}_0$. The vector $\bm{\theta}_0$ is therefore the \textit{true} parameter value which corresponds to a possible value in $\bm{\Theta} \subset \real^p$. We let $\bm{\theta}$ denote a \textit{generic} parameter vector, which should therefore not be confused with the true parameter $\bm{\theta}_0$. In order to discuss the estimation of $\bm{\theta}_0$, let us consider the AV or WV which can be computed on the time series $(X_t)$ for different (dyadic) scales of decomposition $J$. For the purpose of this work we will however consider $J \in \mathcal{J}$ scales such that there are \textit{at least} the same number of scales as of parameters. With this in mind, we introduce a class of estimators of $\bm{\theta}_0$ that we define as follows
\begin{equation}
	\hat{\bm{\theta}} := \underset{\bm{\theta} \in \bm{\Theta} }{\argmin} \; 
	\| \mathbf{f}(\hat{\bm{\nu}}) - \mathbf{f}(\bm{\nu}(\bm{\theta}))\|_{\bm{\Omega}}^2,
	\label{eq:f:estimator}
\end{equation}
where $\hat{\bm{\nu}} \in \real_+^J$ and $\bm{\nu}(\bm{\theta}) \in \real_+^J$ denote respectively a suitable estimator of the AV or WV computed on $(X_t)$ and the model-based counterpart (i.e. the AV or WV implied by the assumed model $F_{\bm{\theta}}$). The vector-valued function $\mathbf{f}(\cdot)$ is such that $\mathbf{f}: \, \real_+^J \mapsto \mathcal{G} \subset \real^J$ and is assumed known. Moreover, $\bm{\Omega} \in \real^{J \times J}$ is a positive definite matrix which, if estimated, shall be denoted as $\widehat{\bm{\Omega}}$ (instead of $\bm{\Omega}$) in order to emphasize the stochastic nature of the matrix. Since the AV is a special case of the (Haar) WV (see \cite{flandrin1992wavelet,percival1994long, percival2015wavelet} for details), we choose to call the class of estimators in Eq. (\ref{eq:f:estimator}) as ``Generalized Method of Wavelet Functional Moments'' estimators (GMWFM). The latter is quite general and includes, among others, the Generalized Method of Wavelet Moments (GMWM) proposed in \cite{guerrier2013wavelet} or the Autonomous Regression Method for Allan Variance (ARMAV) of \cite{Jurado19}. Indeed, the GMWM corresponds to the choice $\mathbf{f}(\mathbf{x}) = \mathbf{x}$, while the ARMAV is based on $\mathbf{f}_i(\mathbf{x}_i) = \log_{10}(\mathbf{x}_i), \; i = 1,\ldots, J$, where $\mathbf{f}_i(\cdot)$ and $\mathbf{x}_i$ denote the $i$-th element of $\mathbf{f}(\cdot)$ and $\mathbf{x}$, respectively. In addition, the GMWM and the ARMAV are based on different but relatively similar choices of the matrix $\bm{\Omega}$.

In this paper, we investigate the requirements on the function $\mathbf{f}(\mathbf{x})$ to ensure that the estimator $\hat{\bm{\theta}}$ is consistent and asymptotically normally distributed. Moreover, we discuss whether an optimal choice for $\mathbf{f}(\mathbf{x})$ exists. For this purpose, we need to define a set of assumptions that will be used in order to investigate these properties. Therefore, let us study the first assumption regarding injectivity of the function $\mathbf{g}(\bm{\theta}) := \mathbf{f}(\bm{\nu}(\bm{\theta}))$ which can be found below.

\setcounter{Assumption}{0}
\renewcommand{\theHAssumption}{otherAssumption\theAssumption}
\renewcommand\theAssumption{\Alph{Assumption}}
\begin{Assumption}[Injectivity]
\label{assum:injectiviy}
The functions $\mathbf{f}(\cdot)$ and $\bm{\nu}(\cdot)$ are such that $\mathbf{f}(\cdot)$ is injective in $\real_+^J$ and $\bm{\nu}(\cdot)$ is injective in $\bm{\Theta}$.
\end{Assumption}\vspace{0.25cm}

If this assumption holds, then a direct consequence is that $\mathbf{g}(\bm{\theta})$ is injective in $\bm{\Theta}$. More precisely, the first part of Assumption \ref{assum:injectiviy} is rather mild since the function $\mathbf{f}(\cdot)$ can be chosen in such a way as to respect this condition. However, the second part of the assumption can be challenging to prove. For example, \cite{guerrier2016identifiability} considered the injectivity of the function $\bm{\nu}(\cdot)$ and provide a series of results allowing to verify this property for various classes of latent time series models. The latter demonstrates that the second part of Assumption \ref{assum:injectiviy} would hold for the class of models considered in \cite{guerrier2013wavelet}, with a few exceptions. For example, if the time series contains a drift with parameter $\omega$ it is necessary to assume that the sign of $\omega$ is known (since $\bm{\nu}(\bm{\theta})$ only depends on $\omega^2$). A general strategy to prove whether Assumption \ref{assum:injectiviy} holds for a specific model can be found in \cite{komunjer2012global} (which is also used in \cite{guerrier2016identifiability}) while in the lemma further on we prove the second requirement of Assumption \ref{assum:injectiviy} (i.e. $\bm{\nu}(\cdot)$ is injective in $\bm{\Theta}$) for the general model considered in \cite{Jurado19}. The latter model is a composite model made by the sum of a (1) quantization noise with parameter $Q^2 \in \real_+$, (2) white noise with parameter $\sigma^2 \in \real_+$, (3) bias instability with parameter $B \in \real_+$, (4) random walk with parameter $\gamma^2 \in \real_+$ and (5) drift with parameter $\omega \in \real_+$.

\begin{Lemma}
\label{lem:injectivity}
Let 
\begin{equation*}
    \bm{\theta} := \left[Q^2 \;\;\; \sigma^2 \;\;\; B \;\;\; \gamma^2 \;\;\; \omega \right] \in \bm{\Theta} \subset \real_+^5,
\end{equation*}
and let $c$ be a positive constant. Then, the function 
\begin{equation*}
    \bm{\nu}_j(\bm{\theta}) := c\left(\frac{3Q^2}{2^{2j}} + \frac{\sigma^2}{2^j} + \frac{2\log(2)}{\pi} B^2 + \frac{\gamma^2 2^j}{3} + \omega^2 2^{2j -1}\right),
    \label{eq:wv:theo}
\end{equation*}
is injective in $\bm{\Theta}$. 
\end{Lemma}\vspace{0.25cm}

\begin{Remark}
The positive constant $c$ is simply related to the choice of the AV or (Haar) WV: in the case of the former we have that $c = 1$ while for the Haar WV we have $c = \nicefrac{1}{2}$.\\
\end{Remark}

\noindent \textsc{Proof:}
First we notice that it is sufficient to show that
\begin{equation*}
    \bm{\nu}^*(\bm{\theta}) = \bm{\nu}^*(\bm{\theta}^*),
\end{equation*}
if and only if $\bm{\theta} = \bm{\theta}^*$, where $\bm{\nu}^*(\bm{\theta})$ denotes the first 5 elements of the vector $\bm{\nu}(\bm{\theta})$. Moreover, the function $\bm{\nu}^*(\bm{\theta})$ can be reparametrized as a function of $\bm{\beta}$ defined as
\begin{equation*}
    \bm{\beta} := \left[Q^2 \;\;\; \sigma^2 \;\;\; B^2 \;\;\; \gamma^2 \;\;\; \omega^2 \right],
\end{equation*}
where the only difference with $\bm{\theta}$ is that the elements $B$ and $\omega$ are squared. Since the latter elements are positive (the sign of $\omega$ is known and is assumed positive for this proof without loss of generality), the square function is also injective and this implies that if the WV is injective for their squares, by composition of injective functions it is also injective for the original values. Therefore, it is sufficient to show that
\begin{equation*}
    \bm{\nu}^*(\bm{\beta}) = \bm{\nu}^*(\bm{\beta}^*),
\end{equation*}
if and only if $\bm{\beta} = \bm{\beta}^*$. We start by computing the Jacobian matrix $\mathbf{J}(\bm{\beta})$ which is defined as
\begin{equation*}
\begin{aligned}
\mathbf{J}(\bm{\beta}) &:= \frac{\partial}{\partial \bm{\beta}^T} \; \bm{\nu}^*(\bm{\beta})\\
&= c \begin{bmatrix}
\nicefrac{3}{2}     & \nicefrac{1}{2}   & \nicefrac{2 \log(2)}{\pi}  & \nicefrac{2}{3}  & 2  \\
\nicefrac{3}{16}     & \nicefrac{1}{4}  & \nicefrac{2 \log(2)}{\pi}  & \nicefrac{4}{3}  & 8  \\
\nicefrac{3}{64}     & \nicefrac{1}{8}  & \nicefrac{2 \log(2)}{\pi}  & \nicefrac{8}{3}  & 32  \\
\nicefrac{3}{256}     & \nicefrac{1}{16}  & \nicefrac{2 \log(2)}{\pi}  & \nicefrac{16}{3}  & 128  \\
\nicefrac{3}{1024}  & \nicefrac{1}{32}  & \nicefrac{2 \log(2)}{\pi}  & \nicefrac{32}{3}  & 512  \\
\end{bmatrix}.
\end{aligned}
\end{equation*}
%
Since $\mathbf{J}(\bm{\beta})$ does not depend on $\bm{\beta}$ we let $\mathbf{J} := \mathbf{J}(\bm{\beta})$ which, based on the mean-value theorem, allows us to write
\begin{equation*}
\begin{aligned}
\bm{\nu}^*(\bm{\beta}) - \bm{\nu}^*(\bm{\beta}^*) &= \bm{\nu}^*(\bm{\beta}) - \left[\bm{\nu}^*(\bm{\beta}) + \mathbf{J} \cdot ( \bm{\beta}^* - \bm{\beta})\right]\\
&= \mathbf{J} \cdot (\bm{\beta} - \bm{\beta}^*).
\end{aligned}
\end{equation*}
Since we have that
\begin{equation*}
    \det \left(\mathbf{J}\right) = c^5\frac{84357 \log(2)}{1024 \pi} > 0,
\end{equation*}
the only solution of the equation
\begin{equation*}
    \mathbf{J} \cdot ( \bm{\beta} - \bm{\beta}^*) = \bm{0},
\end{equation*}
is $\bm{\beta}^* = \bm{\beta}$, which concludes the proof. \hfill $\blacksquare$

\vspace{0.25cm}


Having discussed Assumption \ref{assum:injectiviy} which appears to be reasonable to assume in general (given the different cases in which it is verified), we now consider the other set of assumptions that are needed to prove consistency of the estimator $\hat{\bm{\theta}}$.

\begin{Assumption}[Compactness]
\label{assum:compact}
The set $\bm{\Theta}$ is compact.
\end{Assumption}

\begin{Assumption}[Consistency]
\label{assum:consistent}
For all $j \in \left\{1, \ldots, J\right\}$, we have 
\begin{align*}
    | \hat{\bm{\nu}}_j - {\bm{\nu}}_j(\bm{\theta}_0)| = o_{\rm p}(1).
\end{align*}
Moreover, if $\bm{\Omega}$ is estimated by $\widehat{\bm{\Omega}}$ then we have 
\begin{align*}
    ||\widehat{\bm{\Omega}} - \bm{\Omega}||_{S} = o_{\rm p}(1).
\end{align*}
\end{Assumption}

\begin{Assumption}[Continuity]
\label{assum:contuity:f}
The function $\mathbf{g}(\bm{\theta}) :=\mathbf{f}(\bm{\nu}(\bm{\theta}))$ is continuous in $\bm{\Theta}$.
\end{Assumption}\vspace{0.25cm}

Assumption \ref{assum:compact} is a common regularity condition which is typically assumed for most estimation problems or is replaced by other types of constraints. Its main purpose is to ensure that certain quantities that we will consider in the proofs will be bounded in order to ensure convergence. Assumption \ref{assum:consistent} is rather mild and lower-level conditions equivalent to this assumption can, for example, be found in \cite{percival1995estimation} for the WV (as well as in \cite{guerrier2016fast} under weaker conditions) or by combining these results with the work of \cite{percival1994long} who showed the equivalence between the AV and WV. Finally, Assumption \ref{assum:contuity:f} requires the function $\mathbf{f}(\bm{\nu}(\bm{\theta}))$ to be continuous in $\bm{\Theta}$ which is the case when both $\mathbf{f}(\cdot)$ and $\bm{\nu}(\cdot)$ are continuous within their respective composition domains. Since the function $\bm{\nu}(\bm{\theta})$ is continuous in $\bm{\Theta}$ for nearly all models of interest (such as those considered in \cite{guerrier2016identifiability} or the model discussed in \cite{Jurado19}), it is sufficient for $\mathbf{f}(\cdot)$ to be continuous in $\real_+^J$ to satisfy this assumption. Based on these assumptions, we can state the following consistency result.

\begin{Theorem}
\label{thm:consistent}
Under Assumptions \ref{assum:injectiviy} to \ref{assum:contuity:f}, we have that 
\begin{equation*}
    || \hat{\bm{\theta}} - \bm{\theta}_0 ||_2 = o_{\rm p}(1).
\end{equation*}
\end{Theorem}\vspace{0.25cm}

\noindent \textsc{Proof:} Let
\begin{equation*}
    Q(\bm{\theta}) := 	\| \mathbf{g}(\bm{\theta}_0) - \mathbf{g}(\bm{\theta})\|_{\bm{\Omega}}^2,
\end{equation*}
where $\mathbf{g}(\bm{\theta})$ is defined in Assumption \ref{assum:contuity:f}. Then, we have
\begin{equation*}
    Q(\bm{\theta}) \leq || \bm{\Omega} ||_{S} \| \; \mathbf{g}(\bm{\theta}_0) - \mathbf{g}(\bm{\theta})\|_2^2.
\end{equation*}
Therefore, Assumption \ref{assum:injectiviy} implies that $Q(\bm{\theta})$ has a unique minimum in $\bm{\theta} = \bm{\theta}_0$.

Next, Assumption \ref{assum:contuity:f} directly implies the continuity of the function  $Q(\bm{\theta})$ in $\bm{\Theta}$. Moreover, from the continuous mapping theorem together with Assumptions \ref{assum:consistent} and \ref{assum:contuity:f}, we have $\lvert\hat{\mathbf{g}}_j-\mathbf{g}_j(\bm{\theta}_0)\rvert= o_{\rm p}(1)$ for all $j \in \left\{1, \ldots, J\right\}$. Then, following the same strategy as in \cite{guerrier2016fast} (Proposition 3.1.), we obtain
\begin{equation*}
    \sup_{\bm{\theta} \in \bm{\Theta}}\; | \widehat{Q}(\bm{\theta}) - Q(\bm{\theta})| = o_{\rm p}(1),
\end{equation*}
where
\begin{equation*}
    \widehat{Q}(\bm{\theta}) := 	\| \mathbf{f}(\hat{\bm{\nu}}) - \mathbf{f}(\bm{\nu}(\bm{\theta}))\|_{\widehat{\bm{\Omega}}}^2\, .
\end{equation*}
Therefore, Theorem 2.1 of \cite{newey1994large} can be applied to obtain the consistency of $\hat{\bm{\theta}}$ thereby concluding the proof.\hfill $\blacksquare$ 
\vspace{0.25cm}

Theorem \ref{thm:consistent} implies that any GMWFM estimator is consistent under the same conditions needed to ensure the consistency of the GMWM estimator provided that the function $\mathbf{f}(\cdot)$ is both injective (see Assumption \ref{assum:injectiviy}) and continuous (see Assumption \ref{assum:contuity:f}). Therefore, the requirements on the function $\mathbf{f}(\cdot)$ are rather mild but we shall see that this function has a more relevant impact on the asymptotic distribution of the estimator. Before introducing this result, as for the result on consistency, we first state and discuss relevant assumptions.


\begin{Assumption}[Interior and Convex]
\label{assum:int}
The vector $\bm{\theta}_0$ is such that $\bm{\theta}_0 \in \Int(\bm{\Theta})$ and $\bm{\Theta}$ is convex.
\end{Assumption}
\vspace{0.4cm}

%
%

\begin{Assumption}[Function Differentiability]
\label{assum:cov}
The function $\mathbf{f}(\cdot)$ is such that $\mathbf{f} \in \mathcal{C}^1(\real^J_+, \mathcal{G})$ allowing us to define
%
\begin{equation*}
    \mathbf{F}(\bm{\theta}_0) := \left. \frac{\partial}{\partial \mathbf{x}^T} \; \mathbf{f}(\mathbf{x}) \right|_{\mathbf{x} = \bm{\nu}(\bm{\theta}_0)}\, .
\end{equation*}
Moreover, defining the matrices
\begin{equation*}
    \bm{\Omega}^*\left[\bm{\theta}_0, \mathbf{F}\right] := \mathbf{F}(\bm{\theta}_0)^T \boxtimes  \bm{\Omega},
\end{equation*}
and
\begin{equation*}
    \mathbf{A}(\bm{\theta}_0) := \left. \frac{\partial}{\partial \bm{\theta}^T} \; \bm{\nu}(\bm{\theta}) \right|_{\bm{\theta} = \bm{\theta}_0}\,,
\end{equation*}
then, the matrix
\begin{equation*}
    \mathbf{H}\left[\bm{\theta}_0, \bm{\nu}, \bm{\Omega}, \mathbf{F}\right] := \mathbf{A}(\bm{\theta}_0)^T \boxtimes \bm{\Omega}^*\left[\bm{\theta}_0, \mathbf{F}\right],
\end{equation*}
exists and is non-singular. 


\end{Assumption}


\begin{Assumption}
\label{assum:nu:norm}
The estimator $\hat{\bm{\nu}}$ has the following asymptotic distribution
\begin{equation*}
	\sqrt{T_J}\left(\hat{\bm{\nu}}-\bm{\nu}(\bm{\theta}_0)\right) \xrightarrow[T\rightarrow\infty]{\mathcal{D}} \mathcal{N}\left(\mathbf{0},\mathbf{V}(\bm{\theta}_0)\right),
\end{equation*}	
where $\mathbf{V}(\bm{\theta}_0) := \cov(\hat{\bm{\nu}})$ is a positive-definite symmetric matrix.
\end{Assumption}\vspace{0.25cm}

The topological requirements of Assumption \ref{assum:int} are quite mild although stronger than necessary. Indeed, the fact that $\bm{\theta}_0$ is required to be an interior point of the convex space $\bm{\Theta}$ is convenient (but not strictly necessary) to ensure that expansions (such as Taylor expansions) can be made between $\bm{\theta}_0$ and an arbitrary point in $\bm{\Theta}$. Assumption \ref{assum:cov} contains different requirements but what it basically requires is that the function $\mathbf{f}(\cdot)$ is differentiable in such a way that it can be used to make Taylor expansions for the purposes of demonstrating the asymptotic normality of the estimator $\hat{\bm{\theta}}$. Based on these expansions we obtain expressions that deliver the matrix $\mathbf{H}\left[\bm{\theta}_0, \bm{\nu}, \bm{\Omega}, \mathbf{F}\right]$ which needs to be positive-definite in order for the estimator to have an asymptotic variance (and hence define an asymptotic distribution). Finally, Assumption \ref{assum:nu:norm} is required for any estimator which makes use of moments (such as the AV or WV) to deliver asymptotic normality of the estimator itself. This assumption is verified under few additional conditions compared to those required for Assumption \ref{assum:consistent}, as highlighted again in \cite{percival1995estimation}, \cite{serroukh2000statistical} and, under weaker conditions, in \cite{guerrier2016fast}. Using these assumptions, we obtain the following result.
\begin{Theorem}
\label{thm:asym:norm}
	Under Assumptions \ref{assum:injectiviy} to \ref{assum:nu:norm}, the estimator $\hat{\bm{\theta}}$ has the following asymptotic distribution
\begin{equation*}
	\sqrt{T_J}\left(\hat{\bm{\theta}}-\bm{\theta}_0\right) \xrightarrow[T\rightarrow\infty]{\mathcal{D}} \mathcal{N}\left(\mathbf{0}, \bm{\Sigma}\left[\bm{\theta}_0, \bm{\nu}, \bm{\Omega}, \mathbf{F}\right]\right),
\end{equation*}		
where
\begin{equation*}
	\bm{\Sigma}\left[\bm{\theta}_0, \bm{\nu}, \bm{\Omega}, \mathbf{F}\right] := \mathbf{B}\left[\bm{\theta}_0, \bm{\nu}, \bm{\Omega}, \mathbf{F}\right] \boxtimes \mathbf{V}(\bm{\theta}_0),
\end{equation*}		
and
\begin{equation*}
    \mathbf{B}\left[\bm{\theta}_0, \bm{\nu}, \bm{\Omega}, \mathbf{F}\right] := \mathbf{H}\left[\bm{\theta}_0, \bm{\nu}, \bm{\Omega}, \mathbf{F}\right]^{-1} \mathbf{A}(\bm{\theta}_0)^T \bm{\Omega}^*\left[\bm{\theta}_0, \mathbf{F}\right].
\end{equation*}		
%
\end{Theorem}\vspace{0.25cm}

\noindent \textsc{Proof: } Let $\bm{\Theta}(T) := \left\{\mathbf{x} \in \real^p \,| \,\, ||\mathbf{x} - \bm{\theta}_0 ||_2 \leq d(T) \right\}$, where $d(T) = o(1)$. Moreover, we also define $\bm{\Theta}^*(T) := \bm{\Theta} \cap \bm{\Theta}(T)$. Since $\hat{\bm{\theta}}$ is consistent by Theorem~\ref{thm:consistent} (based on  Assumptions~\ref{assum:injectiviy} to \ref{assum:contuity:f}), there exists a function $d(T)$ such that
\begin{equation}
    \begin{aligned}
    \hat{\bm{\theta}} :=& \underset{\bm{\theta} \in \bm{\Theta} }{\argmin} \; 
	\| \mathbf{f}(\hat{\bm{\nu}}) - \mathbf{f}(\bm{\nu}(\bm{\theta}))\|_{\bm{\Omega}}^2\\
	=&	\underset{\bm{\theta} \in \bm{\Theta}^*(T)}{\argmin} \; 
	\| \mathbf{f}(\hat{\bm{\nu}}) - \mathbf{f}(\bm{\nu}(\bm{\theta}))\|_{\bm{\Omega}}^2 + o_{\rm p}(1).
	 \end{aligned}
	\label{eq:proof:as:norm}
\end{equation}
Within the set $\bm{\Theta}^*(T)$, which shrinks towards $\bm{\theta}_0$ as the sample size $T$ increases, we can expand $\mathbf{f}\left(\hat{\bm{\nu}}\right)$ and $\mathbf{f}\left(\bm{\nu}(\bm{\theta})\right)$ around $\bm{\nu}(\bm{\theta}_0)$ using a Taylor expansion to obtain:
\begin{equation*}
\begin{aligned}
\mathbf{f}\left(\hat{\bm{\nu}}\right) &= \mathbf{f}\left(\bm{\nu}(\bm{\theta}_0)\right) + \mathbf{F}\left(\bm{\theta}_0\right) \left(\hat{\bm{\nu}} - \bm{\nu}(\bm{\theta}_0) \right) + o_{\rm p}(1)\\[0.2cm]
\mathbf{f}\left(\bm{\nu}(\bm{\theta})\right) &= \mathbf{f}\left(\bm{\nu}(\bm{\theta}_0)\right) + \mathbf{F}\left(\bm{\theta}_0\right) \left(\bm{\nu}(\bm{\theta}) - \bm{\nu}(\bm{\theta}_0) \right) + o(1).
\end{aligned}
\end{equation*}
Therefore, by combining this result with (\ref{eq:proof:as:norm}), we obtain
\begin{equation}
    \begin{aligned}
    \hat{\bm{\theta}} &= \underset{\bm{\theta} \in \bm{\Theta}^*(T)}{\argmin} \; 
	\| \mathbf{F}\left(\bm{\theta}_0\right) \left(\hat{\bm{\nu}} - \bm{\nu}(\bm{\theta})\right) \|_{\bm{\Omega}}^2 + o_{\rm p}(1) \\
	&= \underset{\bm{\theta} \in \bm{\Theta}^*(T)}{\argmin} \; 
	\|  \hat{\bm{\nu}} - \bm{\nu}(\bm{\theta}) \|_{\bm{\Omega}^*}^2 + o_{\rm p}(1),
    \end{aligned}
    \label{eq:asymp:equivalence}
\end{equation}
where, similarly to the definition of Assumption \ref{assum:cov}, 
\begin{equation*}
    \bm{\Omega}^* := \bm{\Omega}^*\left[\bm{\theta}_0, \mathbf{F}\right] = \mathbf{F}\left(\bm{\theta}_0\right)^T \boxtimes \bm{\Omega}.
\end{equation*}
Next, we consider the following approximation of $\hat{\bm{\theta}}$,
\begin{equation*}
    \tilde{\bm{\theta}} := \underset{\bm{\theta} \in \bm{\Theta}}{\argmin} \; 
	\|  \hat{\bm{\nu}} - \bm{\nu}(\bm{\theta}) \|_{\bm{\Omega}^*}^2.
\end{equation*} 
Proposition 4.2 of \cite{guerrier2016fast} implies, under the current assumption framework, that
\begin{equation*}
	\sqrt{T_J}\left(\tilde{\bm{\theta}}-\bm{\theta}_0\right) \xrightarrow[T\rightarrow\infty]{\mathcal{D}} \mathcal{N}\left(\mathbf{0},\bm{\Sigma}\left[\bm{\theta}_0, \bm{\nu}, \bm{\Omega}, \mathbf{F}\right]\right).
\end{equation*}	
Since $\hat{\bm{\theta}} = \tilde{\bm{\theta}} + o_{\rm p}(1)$, a direct application of Slutsky's theorem allows to conclude that the above results remains true for $\hat{\bm{\theta}}$ and we obtain
\begin{equation*}
	\sqrt{T_J}\left(\hat{\bm{\theta}}-\bm{\theta}_0\right) \xrightarrow[T\rightarrow\infty]{\mathcal{D}} \mathcal{N}\left(\mathbf{0},\bm{\Sigma}\left[\bm{\theta}_0, \bm{\nu}, \bm{\Omega}, \mathbf{F}\right]\right),
\end{equation*}	
which concludes the proof.\hfill $\blacksquare$
\vspace{0.25cm}

An implication of this result (made evident in particular from Eq. (\ref{eq:asymp:equivalence}) in the proof) is the fact that, no matter which choice is made for the function $\mathbf{f}(\cdot)$ and the matrix $\bm{\Omega}$ (provided that they satisfy the previously mentioned assumptions), we can define a matrix $\bm{\Omega}^*$ (that depends upon $\mathbf{f}(\cdot)$ and $\bm{\Omega}$) such that we can express the estimator as
\begin{equation}
\label{eq:argmin_eq}
    \hat{\bm{\theta}} := \underset{\bm{\theta} \in \bm{\Theta}}{\argmin} \; 
	\|  \hat{\bm{\nu}} - \bm{\nu}(\bm{\theta}) \|_{\bm{\Omega}^*}^2 .
\end{equation}
Therefore, as long as the matrix $\bm{\Omega}^* := \bm{\Omega}^*\left[\bm{\theta}_0, \mathbf{F}\right]$ is positive definite, the estimator $\hat{\bm{\theta}}$ is asymptotically normally distributed under the above assumptions and the only aspect that is affected by the change of $\bm{\Omega}^*$ is the efficiency of the resulting estimator. Consequently, the choice of a specific function $\mathbf{f}(\cdot)$ (which respects the required properties)  only contributes to modifying the weighting matrix $\bm{\Omega}^*$ thereby delivering approximately the same results for any such function $\mathbf{f}(\cdot)$. The weighting matrix $\bm{\Omega}^*$ is therefore crucial to the efficiency of the estimator $\hat{\bm{\theta}}$. As shown in the corollary below, the optimal choice (in terms of asymptotic efficiency) of $\bm{\Omega}^*$ is the inverse of $\mathbf{F}(\bm{\theta}_0) \boxtimes \mathbf{V}(\bm{\theta}_0)$. Although the true $\mathbf{V}(\bm{\theta}_0)$ is unknown in practice, it can be consistently estimated by the estimator proposed in \cite{andrews1991heteroskedasticity} or with the approach discussed in \cite{guerrier2016fast}. 
Moreover, the corollary of Theorem \ref{thm:asym:norm} presented below shows how asymptotically optimal estimators can be constructed for the GMWFM. 

\begin{Corollary}
\label{coro:optim}
Under Assumptions \ref{assum:injectiviy} to \ref{assum:nu:norm} (i.e. the same conditions of Theorem~\ref{thm:asym:norm}), the estimator $\hat{\bm{\theta}}$ based on the function $\mathbf{f}(\cdot)$ and the matrix $\bm{\Omega}^\circ := [\mathbf{F}(\bm{\theta}_0) \boxtimes \mathbf{V}(\bm{\theta}_0)]^{-1}$ is asymptotically efficient in the class of GMWFM estimators.
\end{Corollary}\vspace{0.25cm}


\noindent \textsc{Proof: } 
Under our assumptions, it is easy to verify that the asymptotic covariance matrix of $\hat{\bm{\theta}}$ is given by
\begin{equation*}
    \bm{\Sigma}[\bm{\theta}_0,\bm{\nu},\bm{\Omega}^\circ,\mathbf{F}] = \left[\mathbf{A}(\bm{\theta}_0)^T\boxtimes\mathbf{V}(\bm{\theta}_0)^{-1}\right]^{-1}.
\end{equation*}
We proceed by demonstrating that the difference between the asymptotic covariance matrix in Theorem~\ref{thm:asym:norm} and the above covariance matrix leads to a positive semi-definite matrix. Following Section 5.2 in \cite{newey1994large}, it is easy to show that
\begin{align*}
    \bm{\Sigma}[\bm{\theta}_0,\bm{\nu},\bm{\Omega},\mathbf{F}] &- \bm{\Sigma}[\bm{\theta}_0,\bm{\nu},\bm{\Omega}^\circ,\mathbf{F}]\\
    &=\mathbf{H}\left[\bm{\theta}_0, \bm{\nu}, \bm{\Omega}, \mathbf{F}\right]^{-1}\boxtimes\mathbb{E}\left[\mathbf{W}\mathbf{W}^T\right],
\end{align*}
where 
\begin{align*}
    \mathbf{W} :=
    &\mathbf{A}(\bm{\theta}_0)^T \bm{\Omega}^*\left[\bm{\theta}_0, \mathbf{F}\right]\mathbf{Z}\\ 
    - &\mathbf{H}\left[\bm{\theta}_0, \bm{\nu}, \bm{\Omega}, \mathbf{F}\right] \bm{\Sigma}[\bm{\theta}_0,\bm{\nu},\bm{\Omega}^\circ,\mathbf{F}] \mathbf{A}(\bm{\theta}_0)^T\mathbf{F}(\bm{\theta}_0)^T
    \bm{\Omega}^\circ \mathbf{Z},
\end{align*}
and $\mathbf{Z}$ is a random vector such that
$$\mathbb{E}\left[\mathbf{Z}\mathbf{Z}^T\right] = \mathbf{F}(\bm{\theta}_0) \mathbf{V}(\bm{\theta}_0) \mathbf{F}(\bm{\theta}_0)^T = \left(\bm{\Omega}^{\circ}\right)^{-1}.$$
The result follows since $\mathbb{E}\left[\mathbf{W}\mathbf{W}^T\right]$ is positive semi-definite, which concludes the proof. \hfill $\blacksquare$
\vspace{0.25cm}

Corollary \ref{coro:optim} shows that, under suitable conditions, any estimator belonging to the class of GMWFM estimators can be asymptotically optimal provided that it is based on the function $\mathbf{f}(\cdot)$ and the matrix $\bm{\Omega}^\circ := [\mathbf{F}(\bm{\theta}_0)\mathbf{V}(\bm{\theta}_0) \mathbf{F}(\bm{\theta}_0)^T]^{-1}$. This implies that there exist an infinite number of possible efficient estimators (based on different functions $\mathbf{f}(\cdot)$ and matrix $\bm{\Omega}^\circ$) leading to the same optimal asymptotic covariance matrix $[\mathbf{A}(\bm{\theta}_0)^T\boxtimes\mathbf{V}(\bm{\theta}_0)^{-1}]^{-1}$. In the case where $\mathbf{f}(\mathbf{x}) = \mathbf{x}$, the matrix $\bm{\Omega}^\circ$ has the simplest expression given by $\bm{\Omega}^\circ = \mathbf{V}(\bm{\theta}_0)^{-1}$ since $\mathbf{F}(\bm{\theta}_0) = \mathbf{I}$ thereby also suggesting that its (consistent) estimation is more straightforward in practice. The choice of the function $\mathbf{f}(\mathbf{x}) = \mathbf{x}$ presents several other advantages compared to possible alternative choices. For example, this function allows the estimator to be solved analytically for various commonly used models. This is of particular importance for inertial sensor calibration as most models considered in this field allow for such a closed form solution. Indeed, suppose that there exists a matrix $\mathbf{X}$ that does not depend on $\bm{\theta}$ such that $\bm{\nu}(\bm{\theta})$ can be expressed as $\bm{\nu}(\bm{\theta}) = \mathbf{X}\, \mathbf{h}(\bm{\theta})$ for all $\bm{\theta} \in \bm{\Theta}$, where $\mathbf{h}(\cdot)$ is an injective vector-valued function such that $\mathbf{h}: \, \bm{\Theta} \mapsto \mathcal{H} \subset \real^p$. This is, for example, the case for the model considered in Lemma~\ref{lem:injectivity}. Indeed, denoting $a:=\nicefrac{2\log(2)}{\pi}$, the function $\bm{\nu}(\bm{\theta})$ can be expressed as follows
%
%
\begin{equation}
        \bm{\nu}(\bm{\theta}) =  c \underbrace{\begin{bmatrix}
\frac{3}{2^2} & \frac{1}{2^1} & a & \frac{2^1}{3} & 2^1\\
\frac{3}{2^4} & \frac{1}{2^2} & a & \frac{2^2}{3} & 2^3\\
\vdots & \vdots & \vdots & \vdots & \vdots\\
\frac{3}{2^{2J}} & \frac{1}{2^{J}} & a & \frac{2^J}{3} & 2^{2J-1}
\end{bmatrix}}_{\mathbf{X}} \underbrace{\begin{bmatrix}
Q^2\\
\sigma^2\\
B^2\\
\gamma^2\\
\omega^2
\end{bmatrix}}_{\mathbf{h}(\bm{\theta})} .
\label{eq:example:linear:nu}
\end{equation}
Hence, taking the parameter vector defined in Lemma~\ref{lem:injectivity}, the vector-valued function $\mathbf{h}(\bm{\theta})$ is the identity for all elements except for $B$ and $\omega$ for which it is the square function. Since all parameters (are assumed to) belong to $\real_+$, we have that the function $\mathbf{h}(\bm{\theta})$ is injective which, in general, allows us to write
\begin{equation*}
	\hat{\bm{\theta}} := \underset{\bm{\theta} \in \bm{\Theta} }{\argmin} \; 
	\| \hat{\bm{\nu}} - \bm{\nu}(\bm{\theta})\|_{\bm{\Omega}}^2 = \mathbf{h}^{-1}(\hat{\bm{\vartheta}}),
\end{equation*}
where
\begin{equation*}
	\hat{\bm{\vartheta}} := \underset{\bm{\vartheta} \in \mathcal{H}}{\argmin} \; 
	\| \hat{\bm{\nu}} - \mathbf{X} \, \bm{\vartheta}\|_{\bm{\Omega}}^2,
\end{equation*}
and where we let $\bm{\vartheta} := \mathbf{h}(\bm{\theta})$. Moreover, since the function $\lVert\hat{\bm{\nu}} - \mathbf{X} \, \bm{\vartheta}\rVert_{\bm{\Omega}}^2$ is differentiable in $\bm{\vartheta}$, we have 
\begin{equation*}
	\hat{\bm{\vartheta}} := \underset{\bm{\vartheta} \in \mathcal{H}}{\argmin} \; 
	\| \hat{\bm{\nu}} - \mathbf{X} \, \bm{\vartheta}\|_{\bm{\Omega}}^2 = \underset{\bm{\vartheta} \in \mathcal{H} }{\argzero} \; \mathbf{X}^T \bm{\Omega} \left(\hat{\bm{\nu}} - \mathbf{X} \bm{\vartheta}\right).
\end{equation*}
Therefore, we have that
\begin{equation*}
	  \mathbf{X}^T \bm{\Omega} \mathbf{X} \hat{\bm{\vartheta}} = \mathbf{X}^T \bm{\Omega} \hat{\bm{\vartheta}} \hat{\bm{\nu}},
\end{equation*}
which corresponds to the standard (weighted) least-squares equations. Under Assumption \ref{assum:cov}, $\mathbf{X}^T \bm{\Omega} \mathbf{X}$ is non-singular, we thus obtain
\begin{equation}
	  \hat{\bm{\vartheta}} = \left(\mathbf{X}^T \bm{\Omega} \mathbf{X} \right)^{-1}\mathbf{X}^T \bm{\Omega} \hat{\bm{\nu}},
	  \label{eq:gmwm:simple:trans1}
\end{equation}
which provides a simple plug-in estimator for $\bm{\theta}_0$ defined as
\begin{equation}
	  \hat{\bm{\theta}} = \mathbf{h}^{-1} \left[ \left(\mathbf{X}^T \bm{\Omega} \mathbf{X} \right)^{-1}\mathbf{X}^T \bm{\Omega} \hat{\bm{\nu}} \right].
	  \label{eq:gmwm:simple:trans2}
\end{equation}
%
The above closed-form solution is therefore a first advantage of choosing $\mathbf{f}(\mathbf{x}) = \mathbf{x}$. Moreover there are a few practical advantages stemming from this setting, the first of which is the fact that, given a closed form solution for this class of models, no optimization is required to compute the estimates thereby delivering computationally fast solutions. In addition, even if the model of interest contains a subset of this class of models, this closed form solution can be used as an approximate method to quickly obtain ``good'' starting values that can increase the computational efficiency of the optimization procedure required to solve Eq. \eqref{eq:f:estimator}. Finally, the above form allows to obtain the exact form of the asymptotic variance of $\hat{\bm{\vartheta}}$ up to the value of $\mathbf{V}(\bm{\theta}_0)$ (i.e. the asymptotic variance of $\hat{\bm{\nu}}$) which, using the delta method, would allow us to obtain the exact variance of $\hat{\bm{\theta}}$ for this class of models. 
\begin{figure*}[!h]
  \centering
  \includegraphics[width=1\textwidth]{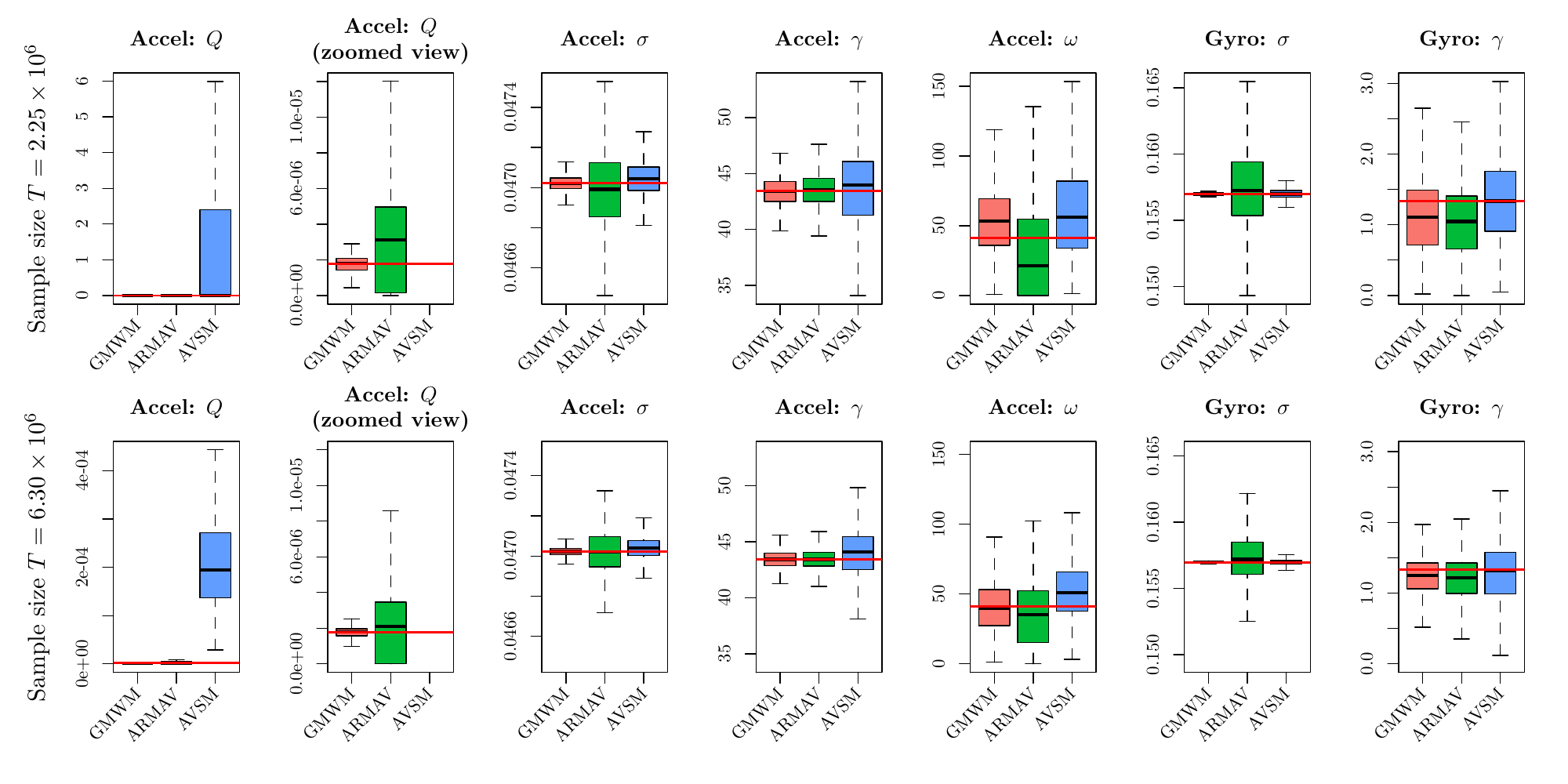}
  \caption{Empirical distribution of the estimations of the GMWM, ARMAV and AVSM approaches for the parameters of the stochastic error of an accelerometer and a gyroscope for signal lengths of $T=\num{2.5e6}$ (top row) and $T=\num{6.3e6}$ (bottom row) respectively.}
  \label{fig:boxplot}
\end{figure*}

Aside from the computational advantage of using the function $\mathbf{f}(\mathbf{x}) = \mathbf{x}$ in Eq. (\ref{eq:f:estimator}) for the class of models for which $\bm{\nu}(\bm{\theta})$ can be expressed as $\mathbf{X}\mathbf{h}(\bm{\theta})$ (or as a ``good'' starting value for other models), there is another potential advantage of using the identity function for the purposes of estimation which relates to their bias. Indeed, the standard estimators of AV or WV are unbiased (see for example~\cite{percival1995estimation}), meaning that $\mathbb{E}\left[\hat{\bm{\nu}}\right] = \bm{\nu}\left(\bm{\theta}_0\right)$ and implying that $\mathbb{E}[ \hat{\bm{\vartheta}} ] = \mathbf{h}(\bm{\theta}_0)$ based on Eq.~\eqref{eq:gmwm:simple:trans1}. A first implication of these properties is that, aside from consistency, it is possible to show that the estimates of most of the parameters in models whose theoretical WV can be expressed as $\mathbf{X}\mathbf{h}(\bm{\theta})$ are unbiased (i.e. when $\mathbf{h}_i(\cdot)$ is the identity or a linear function). However, if the theoretical WV cannot be expressed in the latter form, formal proofs to determine the finite sample behaviour of the estimators defined in Eq. (\ref{eq:f:estimator}) may be hard to derive. Nevertheless, an intuitive argument would support the employment of the function $\mathbf{f}(\mathbf{x}) = \mathbf{x}$ since it directly makes use of unbiased estimators of the WV to match their theoretical counterpart (which is a desirable property in order to achieve unbiasedness with respect to the parameter of interest $\bm{\theta}_0$).


To better highlight the concepts behind the above reasoning, we compare the following two asymptotically equivalent estimators:
\begin{enumerate}
    \item $\hat{\bm{\theta}}$ based on the choice of function $\mathbf{f}(\mathbf{x}) = \mathbf{x}$ and a (non-random) weight matrix $\bm{\Omega}_1$,
    \item $\tilde{\bm{\theta}}$ based on another choice of function, such that $\mathbb{E}\left[\mathbf{f} \left(\hat{\bm{\nu}}\right)\right] \neq \mathbf{f}(\mathbb{E}\left[ \hat{\bm{\nu}}\right])$ and a (non-random) weight matrix $\bm{\Omega}_2$.
\end{enumerate}
We then consider the expected value of the objective function of the first estimator $\hat{\bm{\theta}}$:
\begin{equation*}
    \mathbb{E}\left[\| \hat{\bm{\nu}} - \bm{\nu}(\bm{\theta})\|_{\bm{\Omega}_1}^2\right] = \| \mathbb{E}\left[ \hat{\bm{\nu}} - \bm{\nu}(\bm{\theta})\right]\|_{\bm{\Omega}_1}^2 + \tr \left( \bm{\Omega}_1 \var \left(\hat{\bm{\nu}} \right) \right).
\end{equation*}
Since the second term of the above equation does not depend on $\bm{\theta}$, we let $b_1 := \tr \left( \bm{\Omega}_1 \var \left(\hat{\bm{\nu}} \right) \right)$ and we can write
\begin{equation*}
\begin{aligned}
    \mathbb{E}\left[\| \hat{\bm{\nu}} - \bm{\nu}(\bm{\theta})\|_{\bm{\Omega}_1}^2\right]
    &= \| \bm{\nu}(\bm{\theta}_0) - \bm{\nu}(\bm{\theta})\|_{\bm{\Omega}_1}^2 + b_1,
\end{aligned}
\end{equation*}
since $\mathbb{E}\left[ \hat{\bm{\nu}} \right] =  \bm{\nu}(\bm{\theta}_0)$. Therefore, this function is unbiased in the sense that it is minimized at the true value $\bm{\theta}_0$. 

Following the argument in \cite{han2006gmm}, it is therefore expected that the bias of this estimator will be large if bias in the objective function of the corresponding estimator is large. Next, we consider the objective function of the second estimator and, recalling that $\mathbf{g}(\bm{\theta}) :=\mathbf{f}(\bm{\nu}(\bm{\theta}))$ as defined in Assumption \ref{assum:contuity:f}, we define $\Delta(\bm{\theta}_0) := \mathbb{E}[\mathbf{f}(\hat{\bm{\nu}})] - \mathbf{g}(\bm{\theta}_0)$ and $b_2 := \tr \left( \bm{\Omega}_2 \var \left(\mathbf{f}(\hat{\bm{\nu}}) \right) \right)$. Using these definitions, we obtain
\begin{equation*}
\begin{aligned}
    &\mathbb{E}\left[\| \mathbf{f}(\hat{\bm{\nu}}) - \mathbf{g}(\bm{\theta})\|_{\bm{\Omega}_2}^2\right] = \| \mathbb{E}\left[\mathbf{f}(\hat{\bm{\nu}}) - \mathbf{g}(\bm{\theta}) \right]\|_{\bm{\Omega}_2}^2 + b_2\\
    &\quad =\| \mathbf{g}(\bm{\theta}_0) + \Delta(\bm{\theta}_0) - \mathbf{g}(\bm{\theta}) \|_{\bm{\Omega}_2}^2 + b_2.
\end{aligned}
\end{equation*}
Moreover, by applying the mean value theorem it is possible to assess the order of $\Delta(\bm{\theta}_0)$:
\begin{equation*}
\begin{aligned}
    \Delta(\bm{\theta}_0) &= \mathbb{E}[\mathbf{f}(\hat{\bm{\nu}})] - \mathbf{g}(\bm{\theta}_0)\\
    &=\mathbb{E}[\mathbf{g}(\bm{\theta}_0) + \mathbf{F}\left(\bm{\nu}(\bm{\theta}^*)\right) \left(\hat{\bm{\nu}} - \bm{\nu}(\bm{\theta}_0) \right)] - \mathbf{g}(\bm{\theta}_0)\\
    &= T_J^{-\nicefrac{1}{2}} \; \mathbb{E}[\sqrt{T_J} \mathbf{F}\left(\bm{\nu}(\bm{\theta}^*)\right) \left(\hat{\bm{\nu}} - \bm{\nu}(\bm{\theta}_0) \right)] \\
    &= \mathcal{O}(T_J^{-\nicefrac{1}{2}}),
\end{aligned}
\end{equation*}
where $\bm{\theta}^* \in \bm{\Theta}$, $\bm{\nu}(\bm{\theta}^*)$ is on the line connecting $\bm{\nu}(\bm{\theta}_0)$ and $\hat{\bm{\nu}}$, and the term $\sqrt{T_J} \mathbf{F}\left(\bm{\nu}(\bm{\theta}^*)\right) \left(\hat{\bm{\nu}} - \bm{\nu}(\bm{\theta}_0)\right)$ is $\mathcal{O}_{\rm p}(1)$ by the continuous mapping theorem and Assumption \ref{assum:consistent}. Therefore, we have
\begin{equation*}
\begin{aligned}
    &\underset{\bm{\theta} \in \bm{\Theta} }{\argmin} \; \mathbb{E}\left[\| \mathbf{f}(\hat{\bm{\nu}}) - \mathbf{g}(\bm{\theta})\|_{\bm{\Omega}_2}^2\right]\\
    &\quad = \underset{\bm{\theta} \in \bm{\Theta} }{\argmin} \; \| \mathbf{g}(\bm{\theta}_0) + \Delta(\bm{\theta}_0) - \mathbf{g}(\bm{\theta}) \|_{\bm{\Omega}_2}^2,
\end{aligned}
\end{equation*}
implying that bias of the objective function is of order $\mathcal{O}(T_J^{-1})$ and, consequently, the bias of $\tilde{\bm{\theta}}$ is also of order $\mathcal{O}(T_J^{-1/2})$. As a result, we expect estimators based on the choice $\mathbf{f}(\mathbf{x}) = \mathbf{x}$ to have relatively small biases compared to other choices of $\mathbf{f}(\mathbf{x})$. An important example is when $\mathbf{f}(\mathbf{x})$ is a convex/concave function. More specifically, using Jensen's inequality we have the following
\begin{equation}
\label{eq:jensen_ineq}
    \begin{cases}
        \mathbb{E}\left[\mathbf{f} \left(\hat{\bm{\nu}}\right)\right] > \mathbf{f}(\mathbb{E}\left[ \hat{\bm{\nu}}\right]) & \text{if $\mathbf{f}$ is strictly convex},\\[0.1cm]
        \mathbb{E}\left[\mathbf{f} \left(\hat{\bm{\nu}}\right)\right] < \mathbf{f}(\mathbb{E}\left[ \hat{\bm{\nu}}\right]) & \text{if $\mathbf{f}$ is strictly concave}.\\
    \end{cases}
\end{equation}

Having delivered different theoretical results and arguments comparing the use of different functions $\mathbf{f}(\cdot)$ for the purpose of parameter estimation as defined in Eq. \eqref{eq:f:estimator}, the next section performs some simulation studies where, using different models for the stochastic error of the sensors, we compare the performance of different GMWFM estimators.

\section{Simulation Results}
\label{sec:simulations}

\begin{figure}[!h]
  \centering
  \includegraphics[width=0.49\textwidth]{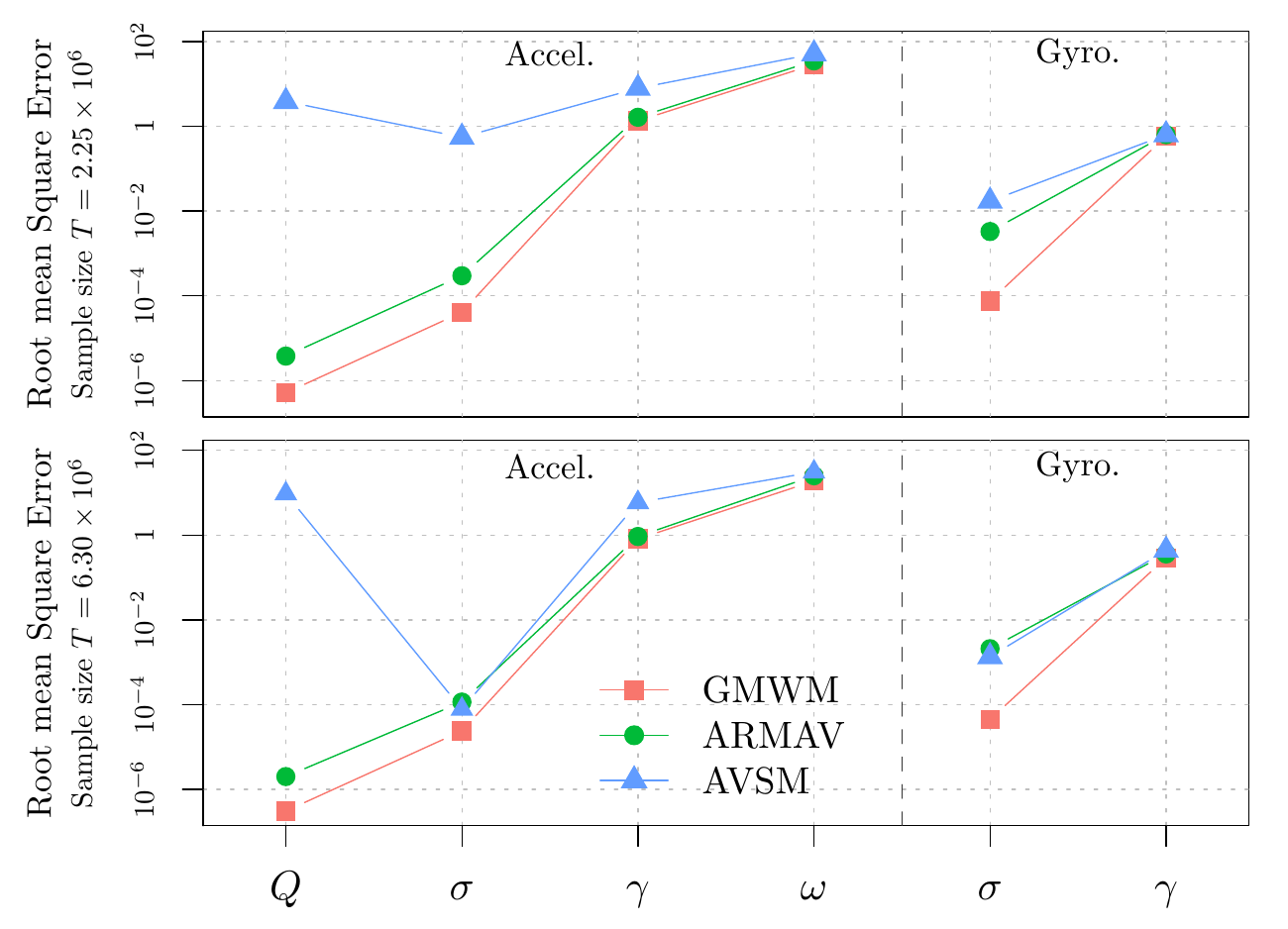}
  \caption{Comparison of estimated root mean squared error betweeen GMWM, ARMAV and AVSM approaches for the parameters of the stochastic error of an accelerometer and a gyroscope for signal lengths of $T=\num{2.5e6}$ (top row) and $T=\num{6.3e6}$ (bottom row) respectively.}
  \label{fig:summary}
\end{figure}

In this section, we compare the estimation performance (in terms of bias and variance) across two GMWFM estimators, namely the GMWM and ARMAV, as well as the standard AVSM as a reference. To carry out this comparison we make use of the parameter estimates (based on the GMWM) of the stochastic processes identified from the real calibration data coming from the X-axis accelerometer and X-axis gyroscope of a STIM-300 IMU \cite{stim}. Hence, the parameter estimates on this real calibration data were considered as being the true parameter values for simulation purposes and their values (along with the respective models) are presented in Table \ref{tab:trueParams}. For each sensor (with respective stochastic models), two Monte Carlo simulation settings were considered based on two different sample sizes (long and short) and with sampling frequency fixed at $250$ Hz. In the ``long" signal setting, the sample size was set to $T=\num{6.3e6}$, corresponding to $7$ hours of calibration data, while in the ``short" signal setting the sample size was set to $T=\num{2.5e6}$, corresponding to $2.5$ hours of calibration data. Each estimation was repeated 3000 times and the empirical distributions of the three estimators are presented in Fig. \ref{fig:boxplot}.

\begin{table}[hbt]
 \centering
  \caption{True parameter values for Monte Carlo simulations}
  \label{tab:trueParams}
  \begin{tabularx}{0.5\textwidth}{*{5}{Y}}
  \toprule
   & \multicolumn{2}{c}{\bf Gyro}  & \multicolumn{2}{c}{\bf Accel}\\
   \cmidrule(lr){2-3} \cmidrule(l){4-5}
   $\bm{\theta}$ & \bf Value & \bf Units & \bf Value & \bf Units\\
   \cmidrule(r){1-1} \cmidrule(lr){2-2} \cmidrule(lr){3-3} \cmidrule(lr){4-4} \cmidrule(l){5-5}
   QN ($Q$) & - & - & \num{1.79e-6} & m/s \\
   WN ($\sigma$) & \num{1.57e-1} & deg/$\sqrt{\text{hr}}$ & \num{4.70e-2} & m/s/$\sqrt{\text{hr}}$ \\
   RW ($\gamma$) & \num{1.34e0} & deg/hr/$\sqrt{\text{hr}}$ & \num{4.35e1} & m/s/hr/$\sqrt{\text{hr}}$ \\
   DR ($\omega$) & - & - & \num{4.14e1} & m/s/hr/hr\\
    \bottomrule
  \end{tabularx}
\end{table}

As shown in Fig. \ref{fig:boxplot}, it would appear that the GMWM approach delivers the best overall performances in terms of bias and dispersion, whereas the ARMAV and AVSM approaches have alternating performances according to the parameter of interest. When comparing the different lengths of the signals (i.e. long and short settings), we can observe that the bias and variance of all GMWFM estimators appear to marginally improve as confirmed in Fig. \ref{fig:summary} that summarizes the overall estimation performance of each approach in terms of Root Mean Square Error (RMSE). From Fig.~\ref{fig:boxplot}-\ref{fig:summary}, the GMWM displays the best performance overall (i.e. lower RMSE) across all parameters regardless of sample length. In contrast, AVSM tends to exhibit larger RMSE, especially when estimating the parameter of the quantization noise ($Q$). In the accelerometer simulations, the ARMAV remains close to the performance of the GMWM while, in the gyroscope simulations, the ARMAV shows similar performance to the AVSM when estimating the white noise parameter ($\sigma$), while all three approaches perform similarly when estimating random walk ($\gamma$).

The simulation results in this section therefore seem to confirm the conclusions made in the previous sections based on the developed theoretical results. Indeed, it would appear that the choice of the function $\mathbf{f}(\mathbf{x}) = \mathbf{x}$ would be the optimal choice when considering a GMWFM estimator for the purpose of (automatic) sensor calibration. 

\section{Conclusions}
\label{sec:conclusion}

This paper discussed the properties and performance of a general class of estimators denoted as GMWFM. Being based on moment-matching techniques through the use of different functions of the WV, these estimators put forward different approaches to perform (automatic) sensor stochastic calibration. Given the variety of proposed functions that build this class of estimators, this paper analysed and proved the properties of such estimators thereby suggesting that the optimal estimator in this class is the one based on the identity function which corresponds to the GMWM. These conclusions are supported by the simulation study which consequently suggest that the GMWM should be the preferred estimator among the GMWFM estimators for the purposes of stochastic calibration of inertial sensors.

\FloatBarrier
\bibliographystyle{unsrt}
\bibliography{ref} 
\end{document}